
\magnification=\magstep 2

\overfullrule=0pt

\hfuzz=16pt

\baselineskip 20pt

\nopagenumbers\pageno=1\footline={\hfil -- {\folio} -- \hfil}

\ 

\centerline{{\bf The Role of Heat Bath and Pointer Modes in Quantum Measurement}}
 
\vskip 0.14 in 

\centerline{Dima Mozyrsky and Vladimir Privman}

\vskip 0.14 in
 
\centerline{\sl Department of Physics, Clarkson University}
\centerline{\sl Potsdam, New York 13699--5820, USA}

\vskip 0.14 in 

\centerline{\bf Abstract} 

We present an exact derivation of a process 
in which a microscopic measured system interacts with ``heat bath''
and pointer modes of a measuring device, via a linear coupling involving
Hermitian operator $\Lambda$ of the system. In the limit of strong interaction 
with these modes, over a small 
time interval, we show that the measured system and the ``pointer'' part of the
measuring device evolve into a statistical mixture of direct-product states
such that the system is in each eigenstate of $\Lambda$ with the correct
quantum-mechanical probability, whereas the expectation values of 
pointer-space operators retain amplified information of the system's
eigenstate.

\vskip 0.14 in


\vfil\eject

The problem of quantum measurement has fascinated scientists
for a long time [1]. It has been argued that a large
``bath'' is an essential ingredient of the measurement
process. Interaction with the bath, which might be a ``heat 
bath'' in thermal equilibrium, causes decoherence which is 
needed to form a statistical mixture of eigenstates out of the
initially fully or partially coherent quantum state of
the measured system. The bath may also play a role
in selection of those quantum states of the pointer
that manifest themselves in classical observations [2-4].

In an exactly solvable model of a quantum oscillator coupled
to a heat bath of oscillators, it has been shown [3] that the
reduced density matrix of the system, with the bath traced over,
decoheres, i.e., it looses its off-diagonal elements in the
eigenbasis of the interaction Hamiltonian. Recent work on
decoherence [5-8] has explored the latter 
effect for rather general cases, for bosonic (oscillator) and 
spin baths. Applications for various physical systems have been
reported [9-15].
Fermionic heat bath has also been used in the literature [16].

It is clear, however, that the full function of a large
measuring device, interacting with a small (microscopic)
quantum system, must be different from thermal 
equilibration or similar averaging effects. The device
must store and amplify
the measurement outcome information. 

The following view of the quantum measurement process
is adopted here. We identify three quantum systems involved. 
First,
the measured system, $S$, is a microscopic system with the
Hamiltonian which will be also denoted by $S$.
Second, the measuring device must
have the ``bath'' or ``body'' part, $B$, containing many individual
modes. The $k$th mode will have the Hamiltonian $B_k$.
The bath part of the device is not observed, i.e., it can be
traced over.
Finally, the device must also have modes
that are not traced over. These modes constitute the pointer, $P$,
that amplifies the information obtained in the measurement
process and can later pass it on for further amplification
or directly to macroscopic (classical) systems.
The $m$th pointer mode has the Hamiltonians $P_m$. It is expected that
expectation values of some quantities in the pointer undergo 
a large change during the measurement process.

It turns out, a posteriori,  that the device modes involved in the 
measurement process can be
quite simple, and they need not interact with
each other. This assumption allows us to focus on the
evolution of the system $S$ and its effect on the pointer $P$.
However, it is the pointer's interaction with the internal
``bath'' or some external modes (the rest of the universe)
that might select those quantum states of $P$
that manifest themselves classically. We avoid the
discussion of this matter, which is in the core of the quantum
measurement ``interpretational'' problem; see [2-4]. Furthermore, 
the measurement process probes
the wavefunction of the measured system at the initial time, $t=0$,
rather than its time evolution under $S$ alone. It is ideally
instantaneous. In practice, it is faster than the time scales
associated with the dynamics under $S$. Such a process
can be obtained as the limit of a system in which very strong
interactions between $S$ and $B$, and also between $S$ 
and $P$, are switched on at $t=0$ and switched off at $t>0$,
with small time interval $t$. At later times, the pointer can 
interact with other systems to pass on
the result of the measurement.

Thus, we assume that the Hamiltonian of the system itself, $S$,
can be ignored in the process. The total Hamiltonian
of the system plus device will be taken 
as
$$ H=\sum_k B_k + \sum_m P_m + b \Lambda 
\sum_k L_k +  p \Lambda \sum_m C_m  \eqno(1) $$
Here
 $\Lambda$ is some Hermitian operator of the system
that couples linearly to certain operators of the modes, $L_k$
and $C_m$. The parameters $b$ and $p$ are
introduced to measure the coupling strength for the
bath and pointer modes, respectively. They are assumed very large;
the ideal measurement process corresponds to $b,p\to \infty$.

We will later specify all the operators in (1) as
the modes of the bosonic heat bath of Caldeira-Leggett
type [14,16-23]. For now, however, let us keep our discussion
general. We will assume that the system operator $\Lambda$ 
has nondegenerate, discrete spectrum of 
eigenstates:
$$ \Lambda | \lambda \rangle = \lambda | \lambda \rangle \eqno(2) $$
Some additional
assumptions on the spectrum of $\Lambda$ and $S$ will be 
encountered later.

Initially, at $t=0$,
the quantum systems ($S,B,P$) and their modes
are not correlated with each other.
We assume that $\rho$ is the initial density matrix of 
the measured system. The initial state of each bath and pointer
mode will be assumed thermalized, with $\beta=1/(kT)$ and
the density matrices
$$ \theta_k = {e^{-\beta B_k}\over{\rm Tr}_k \left(e^{-\beta B_k}\right)}
\qquad\qquad
\sigma_m = {e^{-\beta P_m}\over{\rm Tr}_m \left(e^{-\beta P_m}\right)} \eqno(3) $$
The 
density matrix of the system at time $t$ 
is
$$ R=e^{-iHt/\hbar}\rho\left(\prod_k \theta_k\right)\left( 
\prod_m \sigma_m\right) e^{iHt/\hbar} \eqno(4) $$
The
 bath is not probed and it can be traced over.
The resulting reduced density matrix $r$ 
of the combined system $S+P$ will be
represented as by its matrix elements in the eigenbasis
of $\Lambda$. These
quantities are each an operator in the space 
of $P$:
$$ r_{\lambda\lambda^\prime}=\langle 
\lambda|{\rm Tr}_B(R)|\lambda^\prime\rangle \eqno(5)$$

We now assume that operators in different spaces and of different
modes commute. Then one can show 
that
$$ r_{\lambda\lambda^\prime}=\rho_{\lambda\lambda^\prime} \left[ \prod_m
e^{-i t \left( P_m + p \lambda C_m \right)/\hbar} \sigma_m 
e^{i t \left( P_m + p \lambda^\prime C_m \right)/\hbar} \right] \times $$
$$ \left[ \prod_k {\rm Tr}_k \left\{ e^{-i t \left( B_k + b \lambda L_k \right)/\hbar} \theta_k 
e^{i t \left( B_k + b \lambda^\prime L_k \right)/\hbar} \right\} \right] \eqno(6) $$
where 
$\rho_{\lambda\lambda^\prime}=\langle 
\lambda| \rho |\lambda^\prime\rangle$.
This result involves products of $P$-space operators
and traces over $B$-space operators which are all single-mode. Therefore,
analytical calculations are possible for some choices of the
Hamiltonian (1). The observable $\Lambda$ can be kept general.

The role of the product of traces over the modes of the bath 
in (6) is to induce decoherence
which is recognized as essential for the 
measurement process, e.g., [1,2]. At time $t$, the absolute
value of this product should approach $\delta_{\lambda\lambda^\prime}$
in the limit of large $b$. Let us now assume that the bath is 
bosonic. The Hamiltonian
of each mode is then $ \hbar\omega_k a^\dagger_k a_k$, where
for simplicity we shifted the zero of the oscillator's energy to the
ground state. The coupling operator $L_k$ is usually selected as
$L_k=g^*_ka_k + g_k a_k^\dagger$. For simplicity, though,
we will assume that the coefficients $g_k$ are 
real:
$$B_k = \hbar \omega_k a^\dagger_k a_k
\qquad\quad\quad 
L_k=g_k\left(a_k + a_k^\dagger\right) \eqno(7)$$
For
 example, for radiation field in a unit volume,
coupled to an atom [24],
the coupling is via a linear combination of the operators
$(a_k + a_k^\dagger )/\sqrt{\omega_k}$ and 
$i(a_k - a_k^\dagger )/\sqrt{\omega_k}$. For a spatial oscillator,
these are proportional to position and momentum, respectively. 
Our calculations can be extended to have an
imaginary part of $g_k$ which adds interaction with momentum.

The product of traces in (6) can be calculated by coherent-state
or operator-identity techniques [5-7]. Here and below we only
list the results of such calculations which are usually quite 
cumbersome:
$$  \prod_k {\rm Tr}_k \{ ... \}  = \exp\left\{-2b^2
\left(\lambda-\lambda^\prime\right)^2 \Gamma (t)
+ib^2\left[\lambda^2-(\lambda^\prime)^2\right]\gamma(t) \right\}\eqno(8) 
$$
$$
 \Gamma(t)= \sum_k (\hbar\omega_k)^{-2} g_k^2
\sin^2 {\omega_k t\over 2}
\coth{{\hbar\beta\omega_k \over 2}} \eqno(9) $$
Explicit 
form of $\gamma (t) $ is also known [5].

In the continuum limit of many modes, the density of the bosonic 
bath states in unit volume,
$D(\omega)$, and the Debye cutoff with frequency, $\omega_D$,
are introduced [19] to 
get
$$ \Gamma(t)= \int\limits_0^\infty \, d\omega \,   
{D(\omega) g^2(\omega)\over(\hbar\omega)^2}\, 
e^{-\omega/\omega_D}\, \sin^2 {\omega t\over 2}
\coth{{\hbar\beta\omega \over 2}} \eqno(10) $$
Let us consider the popular choice 
termed Ohmic dissipation [19],
motivated by atomic-physics [24] and
solid-state applications [19], corresponding 
to
$$ D(\omega) g^2(\omega) = \Omega \, \omega \eqno(11)$$
where
$\Omega$ is a constant. Other powers of 
$\omega$ have also been 
considered, e.g., [8].
In studies of decoherence [5-8] for large times $t$, for models without
strong coupling,
not all the choices of $D(\omega) g^2(\omega)$ 
lead to complete decoherence [8] because $\Gamma (t)$
must actually diverge to $+\infty$ for
$t\gg \hbar \beta$, as happens for the choice (11). 

Let us assume
that the largest energy gaps of $S$ are bounded so that there
exists a well defined time scale $\hbar/\Delta S $ of the evolution
of the system under $S$. There is also
the time scale $1/\omega_D$ set by the frequency cutoff assumed
for the interactions. The thermal time scale is $\hbar \beta$. 
The only real limitation on the duration of measurement
is that $t$ must be less then $\hbar/\Delta S$. In applications,
typically [19] one can assume that $1/\omega_D \ll \hbar/\Delta S$.
Furthermore, it is customary to assume that the temperature is 
low [19], 
$$ t {\rm\ and\ } 1/\omega_D \ll \hbar/\Delta S \ll \hbar\beta  \eqno(12) $$
In 
the limit of large $\hbar\beta$, the absolute value of (8) reduces 
to
$$ {\rm Abs} \! \left( \prod_k {\rm Tr}_k \{ ... \} \right) \simeq
 \exp\left\{ -{\Omega
\over 2 \hbar^2} b^2
\left(\lambda-\lambda^\prime\right)^2 \ln [1+(\omega_D t)^2]
\right\}\eqno(13) $$
In 
order to achieve effective decoherence, the product 
$ (\Delta \lambda)^2\, b^2  \ln [1+(\omega_D t)^2] $
must be large. The present 
approach only applies to operators $\Lambda$
with nonzero scale of the smallest spectral gaps, $\Delta \lambda$.

We note that the decoherence property needed for
the measurement process will be obtained for nearly any well-behaved
choice of $D(\omega)g^2(\omega)$ because we can rely on the value
of $b$ being large rather than on the properties of
the function $\Gamma (t)$. If $b$ can be large enough,
very short measurement times are possible. However, it may be advisable
to use measurement times $ 1/\omega_D \ll t \ll \hbar/\Delta S$ to 
get the extra amplification factor $\sim \ln (\omega_D t)$ and allow
for fuller decoherence and less sensitivity to the value of $t$ in the
pointer part of the dynamics, to be addressed shortly.
We notice, furthermore, that the assumption
of a large number of modes is important for monotonic decay
of the absolute value of (8) in decoherence studies [5-8], where irreversibility
is obtained only in the limit of infinite number of modes. In our case, it
can be shown that 
the main role of such a continuum limit is to allow to 
extend the possible measurement times
from $t \ll 1/\omega_D$ to $1/\omega_D \ll t \ll \hbar/\Delta S$.

Consider the reduced density matrix
$r$ of $S+P$, see (6). It becomes diagonal in $|\lambda\rangle$,
at time $t$, because all the nondiagonal elements are  
small,
$$ r=\sum_\lambda |\lambda \rangle \langle \lambda | \, \rho_{\lambda\lambda}
\prod_m e^{-i t \left( P_m + p \lambda C_m \right)/\hbar} \sigma_m 
e^{i t \left( P_m + p \lambda C_m \right)/\hbar} \eqno(14)$$
This result describes a statistically distributed
system, without quantum correlations, 
with the probability $\rho_{\lambda\lambda}$ of the
appropriate  eigenstate $|\lambda \rangle$.

The information in the pointer, perhaps after several steps of amplification,
will be available for probe by interactions with classical devices. 
Foundation of quantum 
mechanics issues that we do not address
include, for instance, the matter of when does the wavefunction
collapse occur for each individual experiment. Is it after time $t$
or after the pointer has been first probed to record
and/or pass to a physicist's brain, the measurement outcome?

At time $t=0$, expectation values of various operators 
of the pointer will have their initial values. These values will
be changed at time $t$ of the measurement owing to the
interaction with the measured system. It is expected that
the large coupling parameter $p$ will yield large 
changes in expectation values of the pointer quantities. 
This does not apply equally to
all operators in the $P$-space. Let us begin with the simplest choice: 
the Hamiltonian $\sum\limits_m P_m$
of the pointer. We will assume that the pointer is described by
the bosonic heat bath and, for simplicity, use the same notation
for the pointer modes as that used for the bath modes. 

The effective density matrix of the pointer for the system's state
$\lambda$ is the product over the $P$-modes in (14). The
expectation value of the pointer energy $E_P$ can be calculated 
from
$$ \langle E_P \rangle_\lambda\,{\rm Tr}_P \left(e^{-\hbar\beta \sum_s 
\omega_s a^\dagger_s a_s }\right)=
{\rm Tr}_P \left\{ \left( \sum_m \hbar\omega_m a^\dagger_m a_m 
\right) \right. \times \qquad\qquad  \eqno(15)
$$
$$
\left. \prod_n \left[ e^{-i t [ \omega_n a^\dagger_n a_n + p \lambda 
g_n(a_n + a_n^\dagger) ]/\hbar} \left(e^{-\hbar\beta \sum_k 
\omega_k a^\dagger_k a_k}\right) e^{i t [ \omega_n a^\dagger_n a_n + p \lambda 
g_n(a_n + a_n^\dagger) ]/\hbar}\right] \right\} $$
where 
we used the thermal initial state (3).
The right-hand side can be reduced to calculations for individual modes.
Operator identities can be then utilized to obtain the 
results
$$ \langle E_P \rangle_\lambda (t) = \langle E_P 
\rangle (0) + \langle \Delta E_P \rangle_\lambda (t)\eqno(16)
$$
$$ 
\langle E_P \rangle (0) = \hbar \sum_m \omega_m e^{-\hbar\beta \omega_m} \left(1-
e^{-\hbar\beta \omega_m}\right)^{-2} \eqno(17) 
$$
$$
\langle \Delta E_P \rangle_\lambda (t)=  {4 p^2 \lambda^2 \over \hbar} 
\sum_m {g_m^2 \over 
\omega_m} \sin^2 \left({\omega_m t \over 2}\right) \eqno(18) $$
For 
a model with Ohmic dissipation, the resulting integral, in the
continuum limit, can be calculated to 
yield
$$ \langle \Delta E_P \rangle_\lambda (t)=  {2\, 
\Omega \,\omega_D \lambda^2 p^2 \over \hbar}
{(\omega_Dt)^2 \over 1 + (\omega_Dt)^2} \eqno(19)$$
which
 should be compared to the exponent in (13). The energy will be an indicator
of the amplified value of the square of 
$\lambda$, provided $p$
is large. Furthermore, we see here the advantage of
larger measurement times, $t \gg 1/\omega_D$. The change in the
energy then reaches saturation. After time $t$, when the
interaction is switched off, the energy of the pointer
will be conserved.

Let us consider the expectation value of the following 
Hermitian operator of the 
pointer:
$$ X=\sum_m C_m = \sum_m g_m (a_m +a_m^\dagger) \eqno(20) $$
For
an atom in a field, $X$ is related to the electromagnetic
field operators [24]. One can show that
$\langle X_P \rangle (0) =0$ 
and
$$ \langle \Delta X_P \rangle_\lambda (t)=  \langle 
X_P \rangle_\lambda (t) = -{4 p \lambda \over \hbar} \sum_m {g_m^2 \over 
\omega_m} \sin^2 \left({\omega_m t \over 2}\right) 
\qquad\qquad \eqno(21) 
$$
$$ 
=  -{2\, 
\Omega \,\omega_D \lambda p \over \hbar}
{(\omega_Dt)^2 \over 1 + (\omega_Dt)^2} $$
The 
change in the expectation value of $X$ is linear in $\lambda$.
However, this operator is not
conserved. One can show that after time $t$ its expectation value 
decays to zero for times $t + {\cal O} (1/\omega_D)$.

We note that by referring to ``unit volume'' we have avoided
the discussion of the ``extensivity'' of various quantities. 
For example, the initial energy $\langle E_P 
\rangle (0)$ is obviously
proportional to the system volume, $V$. However, 
the change 
$\langle \Delta E_P \rangle_\lambda (t)$ 
will not be extensive; typically,
$g^2(\omega)\propto 1/V$, $D(\omega) \propto V$. Thus, while the amplification
in our measurement process can involve a numerically large 
factor, the changes in the quantities
of the pointer will be multiples of microscopic values. Multi-stage
amplification, or huge coupling parameter $p$, would be needed for
the information in the pointer to become truly ``extensive'' macroscopically.

In summary, we described a models of a part of a measurement process, 
involving decoherence
due to a bath and transfer of information 
to a large system (pointer) via
strong interaction over of short period of time.

\vfil\eject

\centerline{\bf References}{\frenchspacing
 
\vskip 0.14 in
 
\item{1.} For a historical overview, see, e.g., A. Whitaker, 
{\it Einstein, Bohr and the Quantum Dilemma}\ (Cambridge 
University Press, 1996).

\item{2.} W. H. Zurek, Physics Today, October 1991, p. 36.

\item{3.} W. G. Unruh, W. H. Zurek, Phys. Rev. D {\bf 40},
1071 (1989).

\item{4.} W. H. Zurek, S. Habib and J. P. Paz,
Phys. Rev. Lett. {\bf 70}, 1187 (1993).

\item{5.} D. Mozyrsky and V. Privman, J. Stat. Phys. {\bf 91},
787 (1998).

\item{6.} N. G. van Kampen, J. Stat. Phys. {\bf 78},
299 (1995).

\item{7.} J. Shao, M.-L. Ge and H. Cheng, Phys. Rev. E {\bf 53},
1243 (1996).

\item{8.} G. M. Palma, K. A. Suominen and A. K. Ekert,
Proc. Royal Soc. London A {\bf 452}, 567 (1996).

\item{9.} I. S. Tupitsyn, N. V. Prokof'ev, P. C. E. Stamp,
Int. J. Modern Phys. B {\bf 11}, 2901 (1997).

\item{10.} C. W. Gardiner {\it Handbook of Stochastic Methods
for Physics, Chemistry and the Natural Sciences}
(Springer-Verlag, Berlin 1990).

\item{11.} A. J. Leggett, in {\it Percolation, Localization and
Superconductivity}, NATO ASI Series B: Physics, Vol. {\bf 109},
edited by A. M. Goldman and S. A. Wolf (Plenum, New York 1984), p. 1.

\item{12.} J. P. Sethna, Phys. Rev. B {\bf 24}, 698 (1981).

\item{13.} Review: A. O. Caldeira and A. J. Leggett,
Ann. Phys. {\bf 149}, 374 (1983).

\item{14.} A. Garg, Phys. Rev. Lett. {\bf 77}, 764 (1996).

\item{15.} L. Mandel and E. Wolf, {\it Optical Coherence 
and Quantum Optics} (Cambridge University Press, 1995).

\item{16.} L.-D. Chang and S. Chakravarty, Phys. Rev. B
{\bf 31}, 154 (1985).

\item{17.} A. O. Caldeira and A. J. Leggett,
Phys. Rev. Lett. {\bf 46}, 211 (1981).

\item{18.} S. Chakravarty and A. J. Leggett, Phys. Rev. Lett.
{\bf 52}, 5 (1984).

\item{19.} Review: A. J. Leggett, S. Chakravarty, A. T. Dorsey,
M. P. A. Fisher and W. Zwerger, Rev. Mod. Phys. {\bf 59}, 1
(1987) [Erratum {\it ibid.\/} {\bf 67}, 725 (1995)].

\item{20.} A. O. Caldeira and A. J. Leggett, Physica {\bf 121A},
587 (1983).

\item{21.} R. P. Feynman and A. R. Hibbs,
{\it Quantum Mechanics and Path Integrals} 
(McGraw-Hill Book Company, 1965).

\item{22.} G. W. Ford, M. Kac and P. Mazur,
J. Math. Phys. {\bf 6}, 504 (1965).

\item{23.} A. J. Bray and M. A. Moore, Phys. Rev. Lett.
{\bf 49}, 1546 (1982).

\item{24.} W. H. Louisell, {\it Quantum Statistical
Properties of Radiation} (Wiley, New York, 1973).

}
 
\bye